\documentclass[twocolumn,showpacs,showkeys,amssymb,nobibnotes,superscriptaddress,aps,pra]{revtex4-1}
\usepackage{amssymb}
\usepackage{amsmath}
\usepackage{subeqnarray}
\usepackage{cases}
\usepackage{amsfonts}
\usepackage{bm}
\usepackage{graphicx}
\usepackage[dvipsnames]{xcolor}
\usepackage{calc}
\usepackage{epsfig}
\usepackage{color}
\usepackage[colorlinks,linkcolor=blue,urlcolor=blue,citecolor=blue]{hyperref}

\begin{document}
\author{Wen Huang}
\affiliation{School of Physics and Hubei Key Laboratory of Gravitation and Quantum Physics, Huazhong University of Science and Technology, Wuhan, 430074, P. R. China}
\affiliation{Wuhan institute of quantum technology, Wuhan, 430074, China}
\author{Ying Wu}
\affiliation{School of Physics and Hubei Key Laboratory of Gravitation and Quantum Physics, Huazhong University of Science and Technology, Wuhan, 430074, P. R. China}
\affiliation{Wuhan institute of quantum technology, Wuhan, 430074, China}
\author{Xin-You L{\"u}}\email{xinyoulu@hust.edu.cn}
\affiliation{School of Physics and Hubei Key Laboratory of Gravitation and Quantum Physics, Huazhong University of Science and Technology, Wuhan, 430074, P. R. China}
\affiliation{Wuhan institute of quantum technology, Wuhan, 430074, China}
\date{\today}

\title{Superradiant phase transition induced by the indirect Rabi interaction}
\begin{abstract}
We theoretically study the superradiant phase transition (SPT) in an indirect Rabi model, where both a two-level system and a single mode bosonic field couple to an auxiliary bosonic field. We find that the indirect spin-field coupling induced by the virtual excitation of the auxiliary field can allow the occurrence of a SPT at a critical point, and the influence of the so-called $A^2$ term in the normal Rabi model is naturally avoided. In the large detuning regime, we present the analytical expression of quantum critical point in terms of the original system parameters. The critical atom-field coupling strength is tunable, which will loosen the conditions on realizing the SPT. Considering a hybrid magnon-cavity-qubit system, we predict the squeezed cat state of magnon generated with feasible experimental parameters, which has potential applications in quantum metrology and quantum information processing.
\end{abstract}
\maketitle
\section{Introduction}\label{Sec I}
\begin{figure*}
\centering
\includegraphics[width=2\columnwidth]{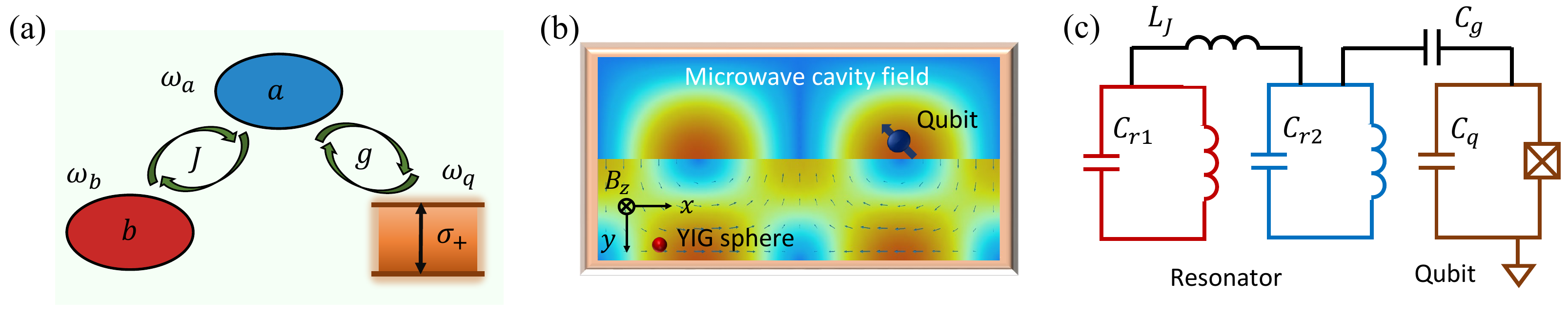}
\caption{\label{fig1} (a) Schematic of an indirect Rabi model. A two-level emitter $\sigma_{+}$ and a bosonic mode $b$ simultaneously interact with an auxiliary mode $a$ with strengths $g$ and $J$, respectively. The frequencies of system are $\omega_q$, $\omega_b$, and $\omega_a$. (b) The realizations of this model in a hybrid magnon-cavity-qubit system, where a superconducting qubit and a YIG sphere couple to the microwave cavity field, simultaneously. A local static magnetic field is applied to make the YIG sphere a single-domain ferromagnet. (c) The realizations of this model in a hybrid circuit QED system which is comprised of two $LC$ resonators and a qubit. The two $LC$ resonators are coupled through inductive coupling and one resonator couples to the qubit through a capacitor.}
\end{figure*}

Quantum phase transition has been the subject of tremendous importance in quantum physics, which not only is fundamentally interesting, but also provides remarkable advantages for quantum techniques \cite{PRE2003chosQPT,PRL2003Entanglement,LambertPRL2004Entanglement,Wang2014NJP,PRA2016DynamicalQM,Lu2018PRA,PRL2020RabiCQM,Quantum2022DMQM}. In the 1970's, the superradiant phase transition \cite{Hepp1973DM,Wang1973DM} was proposed in the Dicke model, which describes $N$ two-level emitters interacting with a single mode bosonic field. In the thermodynamic limit, i.e., $N\rightarrow\infty$, the system suddenly transitions from normal phase to superradiant phase, when tuning the light-matter coupling strength to a critical point. In the superradiant phase, the emitters can emit light in a coherent manner and the radiance intensity proportional to $N^{2}$, i.e., the superradiance, which means the ground state is a superradiative state. Then the process of the SPT can provide a method to realize superradiant effect \cite{PRB2007CollectiveFluorescence,nc2014superradiance}. Typically, this phase transition occurs at zero temperature and thus it is induced by the quantum fluctuations. The widespread attentions have been paid to explore SPT based on the Dicke model in both theoretical \cite{Li2006PRA,ChenGang2007PRADickePhase,PRL2012NonequilibriumQPT1,PRL2014twocoupleDM,PRL2018Dissipation,Lu2018PRApp,Pu2019PRL,Zhu2020PRL,LiuWuMing2020PRL} and experimental \cite{nature2010gas,PRL2014Raman,PRL2016Circuit} studies.

However, the SPT occurs not only in $N\rightarrow\infty$ limit, but also in finite-emitter systems. Recently, it is shown from Ref.\,\cite{PRL2015Hwang} that the quantum Rabi model (a single two-level emitter coupled to a single mode bosonic field) also can undergo a SPT, when the ratio of atomic transition frequency $\Omega$ to the field frequency $\omega$ approaches infinity, i.e., $\Omega/\omega\rightarrow\infty$. Such SPT requires a large light-matter interaction strength located in the ultra-strong coupling regime \cite{nature2009UltrastrongEvolves,np2010ultrastrong,PRL2010SwitchableUltrastrong,PRX2012UltrastrongDynamics,RMP2013Xiang,
PRB2016UltrastrongTwoResonator,np2017SuperconductingUltrastrong,GU2017MicrowavePhotonics,RevModPhys2019Ultrastrong}, which has been probed in a single trapped ion setup \cite{PRL2017TrappedIon,Duanlm2021ion} and nuclear magnetic resonance (NMR) quantum simulator \cite{nc2021NMR} with the great progress of experimental technologies. Motivated by the SPT of the Rabi model, there has been considerable amount of studies concentrating on the extended Rabi models \cite{XiePRX2014Anisotropic,PRA2015twoqubitRabi,PRL2017anisotropic,PRA2018effectiveTwomodeRabi,PRA2018twophotonQRMcircuits,PRA2020twophotonSpectralCollapse,PRA2020asymmetric,PRA2020RabiStarkQPT,
PRA2020MixedLinearNonlinearQRM,PRA2021twomodeQPT,PRL2021MultiqubitMultimodeQRM,PRA2021MixedLinearNonlinearQRM,PRL2021RabiTriangle}. The coexistence of the first-order and second-order quantum phase transition is predicted in the anisotropic Rabi model \,\cite{PRL2017anisotropic}, whose coupling strengths of rotating term and counter-rotating term are anisotropic. Another
expansion is the two-mode Rabi model \cite{PRA2021twomodeQPT} composed of two field modes interacting with a common qubit simultaneously, and it has revealed a new critical point smaller than the normal Rabi model which means the light-matter coupling strength is largely loosened. Apart from these linear expansions, nonlinear Rabi models were proposed to explore various novel physical properties, e.g. two-photon Rabi model \cite{PRA2018twophotonQRMcircuits,PRA2020twophotonSpectralCollapse}, Rabi-Stark model \,\cite{PRA2020RabiStarkQPT}, mixed linear-nonlinear Rabi model \,\cite{PRA2020MixedLinearNonlinearQRM,PRA2021MixedLinearNonlinearQRM}. Indeed, plentiful fascinating phenomenons have been founded in the quantum systems where the two-level system directly couples to a single bosonic mode, such as the new universality classes \cite{PRA2020RabiStarkQPT,PRA2021MixedLinearNonlinearQRM}, spectral collapse \cite{PRA2020twophotonSpectralCollapse}, multicriticalities \cite{PRA2021MixedLinearNonlinearQRM}, as well as rich phase diagram \cite{PRA2021MixedLinearNonlinearQRM,PRL2021RabiTriangle}. However, some direct spin-boson coupling strength is very weak in newly-developing systems \cite{Tabuchi2015science,Tabuchi2017scienceadvances}, e.g., qubit-magnon interaction. A natural question is whether the indirect spin-boson interaction could induce the occurrence of SPT. The crossover between quantum criticality and indirect interaction becomes an important issue, which remains largely unexplored.

Here we propose to realize the SPT based on an indirect Rabi model and discuss the associated quantum criticality. The indirect Rabi model means that a two-level system indirectly couples to a bosonic field via an auxiliary mode, as shown in Fig.\,\ref{fig1}(a). When the frequencies of the two-level system and the field mode are far detuned from the auxiliary mode frequency, an effective Rabi model depending on the original system parameters is obtained after eliminating the auxiliary mode. Using the diagonalization approach and order parameter analysis, we find that the second order quantum phase transition occurs without the requirement of the direct Rabi-type interaction. This SPT is naturally immune to the so-called $A^2$ term appeared in the direct spin-field interactions, and it has a tunable critical coupling strength compared with the standard Rabi model. Our proposal is general, and it can be implemented in a hybrid magnon-cavity-qubit system. By presenting the Wigner function distribution in the phase-space, we clearly show the appearance of the squeezed cat state in the superradiant phase, which offers an alternative method for obtaining a magnon macroscopic quantum superposition. Our work might inspire the following study of the applications of indirect Rabi model in quantum precision measurement.
\section{MODEL AND HAMILTONIAN}\label{sec II}
We consider an indirect Rabi model shown in Fig.\,\ref{fig1}(a), which is applicable to a variety of physical systems, with a total Hamiltonian $(\hbar=1)$
\begin{eqnarray}\label{eq:initial}
H & = & \frac{\omega_{q}}{2}\sigma_{z}+\omega_{a}a^{\dagger}a+\omega_{b}b^{\dagger}b+
g(a^{\dagger}+a)\sigma_{x}\nonumber \\
   &  &
   +J(a^{\dagger}+a)(b^{\dagger}+b),
\end{eqnarray}
where $a$ ($a^{\dagger}$) and $b$ ($b^{\dagger}$) are the annihilation (creation) operators of the bosonic modes with different frequency $\omega_{a}$ and $\omega_{b}$, respectively, and $\sigma_{z}$, $\sigma_{x}$ are the Pauli operators of the two-level emitter with transition frequency $\omega_{q}$. The parameter $g$ describes the coupling strength between the bosonic mode $a$ and the two-level emitter, $J$ denotes the hopping amplitude between the bosonic modes $a$ and $b$. Note that, the uncoupled two-level emitter and bosonic mode $b$ constitute the main part of our model. We use $\!|\!\downarrow\rangle$ and $\!|\!\uparrow\rangle$ to denote the eigenstates of $\sigma_{z}$, and $|n\rangle_{a}$ and $|n\rangle_{b}$ are the eigenstates of $a^{\dagger}a$ and $b^{\dagger}b$, respectively. The parity operator $\Pi=e^{i\pi[a^{\dagger}a+b^{\dagger}b+\frac{1}{2}(1+\sigma_{z})]}$ commutes with the Hamiltonian \,\eqref{eq:initial}, indicating that the system posses a $Z_{\textrm{2}}$ symmetry. Considering the hopping amplitude $J=0$, the Hamiltonian \,\eqref{eq:initial} becomes a standard Rabi model in which a superradiant quantum phase transition occurs when the coupling strength exceeds a critical point, i.e., $g>\sqrt{\omega_{q}\omega_{a}}/2$. Therefore, the bosonic mode $a$ generates a ground-state superradiance. However, we are interested in whether the bosonic mode $b$, without directly interacting with the two-level emitter, can exhibit such a superradiance. Thus, we will focus on the condition of $J\neq0$ and analyze the physical mechanism of SPT in the indirect Rabi model.

Actually, the indirect Rabi model just describes the hybrid magnon-cavity-qubit system depicted in Fig.\ref{fig1}(b), which is experimentally realized in Refs.\,\cite{Tabuchi2015science,Tabuchi2017scienceadvances} based on the quantum magnonics \cite{CRPhys2016Tabuchi,APE2019magnonics}. The collective mode of spins in YIG (yttrium iron garnet) sphere, termed as the magnon mode, has a self-Hamiltonian $-\gamma B_{z}S_{z}$, where $\gamma=28$ GHz/T is the gyromagnetic ratio and $B_{z}$ is the external bias magnetic field. The collective spin operators are defined as $S_{x,y,z}=\sum_{n}s_{x,y,z}^{n}$ and then the raising and lower-ing operators $S_{\pm}=S_{x}\pm iS_{y}$ can be introduced. Using the Holstein-Primakoff transformations \cite{HPtransformation}: $S_{+}=(\sqrt{2S-m^{\dagger}m})m$, $S_{-}=m^{\dagger}(\sqrt{2S-m^{\dagger}m})$, $S_{z}=S-m^{\dagger}m$, the collective spin operators can be expressed in terms of the magnon operator $m$ and $m^{\dagger}$, where $S$ is the total spin number. The self-Hamiltonian is reduced to $\omega_{m}m^{\dagger}m$, where $\omega_{m}=\gamma B_{z}$ is the magnon frequency that can be adjusted from a few hundred $\text{MHz}$ to a few ten $\text{GHz}$. The macrospin is coupled to the microwave photons via magnetic dipole interaction, which can be described by the Hamiltonian $g_{s}(a^{\dagger}+a)(S_{+}+S_{-})$ \cite{magnoncavity}. For low spin excitations ($m^{\dagger}m\ll2S$), one has $S_{+}\approx\sqrt{2S}m$ and $S_{-}\approx\sqrt{2S}m^{\dagger}$. Thus, the interaction becomes $g_{s}\sqrt{2S}(a^{\dagger}+a)(m+m^{\dagger})$, where $g_{s}$ is the coupling strength between the photon and a single spin. The superconducting qubit, acting as a two-level system, is electrically coupled to the cavity field. Note that the direct coupling between the qubit and the magnon mode is too weak, and then it can be ignored safely. Moreover, it was founded that the either the qubit-cavity coupling rate or the magnon-cavity coupling rate can enter into the strong-coupling (SC) regime \cite{prl2011StrongCoupling3DcavityQubit,PRAppl2014HighCooperativity,npjQI2015Strong,PRL2019Strongcoupling} and further into the ultra-strong coupling (USC) regime \cite{PRB2012UltrastrongCoupling3DcavityQubit,PRL2014Ultrastrong,PRB2016UltrahighCooperativity,NJP2019UltrastrongApplications,PRAppl2021Ultrastrong}.

Another platform for performing this model is the hybrid circuit QED system \cite{nc2017HybridQRM}, shown in Fig.\,\ref{fig1}(c). Here, the $LC$ resonator, made of a capacitor and an inductor, takes the place of the bosonic mode, and the transmon qubit serves as the two-level system. The coupling rate between the resonator and qubit has been realized in the USC regime \cite{nc2017HybridQRM,nature2009UltrastrongEvolves,np2010ultrastrong,PRL2010SwitchableUltrastrong,PRX2012UltrastrongDynamics,RMP2013Xiang,
PRB2016UltrastrongTwoResonator,np2017SuperconductingUltrastrong,GU2017MicrowavePhotonics,RevModPhys2019Ultrastrong} or even in the deep-strong coupling (DSC) regime \cite{nc2017DeepStrong}. Analogously, these two resonators are coupled to each other and recently have the USC regime been achieved with recent state-of-the-art technologies, like mediated by a superconducting interference device (SQUID) \cite{PRApp2021UltrastrongResonatorsSQUID} or a Josephson junction \cite{PRL2018UltrastrongTwoOscillators}.

\section{The occurrence of SPT}\label{sec III}
We now assume that the frequency $\omega_{a}$ of the auxiliary mode $a$ is much larger than the frequency $\omega_{b}$ ($\omega_{q}$) of the bosonic mode $b$ (the two-level system), and the detunings $\Delta_{b}=\omega_{a}-\omega_{b}$ and $\Delta_{q}=\omega_{a}-\omega_{q}$ are much larger than the coupling strengths $J$ and $g$, i.e., the large detuning regime. In this case, there is not obvious energy exchange between the ancillary mode and the two-level system (and $b$ mode) \cite{ZhengshibiaoPRL2000,nature2007cavitybus} and the average occupation of the auxiliary mode is very small called virtual excitation. Although the excitation of the auxiliary mode is very small, it establishes a coupling channel between the $b$ mode and the 2-level emitter due to the fact that the auxiliary mode couples to the $b$ mode and the 2-level emitter simultaneously. Correspondingly, the Hamiltonian $H$, after eliminating the degree of freedom of the bosonic mode $a$ with the Fr{\"o}hlich-Nakajima transformation \cite{Frohlich1950,Nakajima1953} $U_{V}=e^{V}$ up to the second order, becomes (see details in the Appendix \,\ref{sec A})
\begin{eqnarray}\label{eq:effective}
H_{\text{eff}} & = & U_{V}^{\dagger}HU_{V}\nonumber \\
 & = & \frac{\omega_{q}}{2}\sigma_{z}+\omega_{b}b^{\dagger}b-
\chi(b^{\dagger}+b)\sigma_{x}-\xi(b^{\dagger}+b)^{2},
\end{eqnarray}
where the generator is
\begin{eqnarray}
V & = & g/\eta_{q}(a\sigma_{-}-a^{\dagger}\sigma_{+})+g/\Delta_{q}(a\sigma_{+}-a^{\dagger}\sigma_{-})\nonumber \\
 &  &
 +J/\eta_{b}(ab-a^{\dagger}b^{\dagger})+J/\Delta_{b}(ab^{\dagger}-a^{\dagger}b).
\end{eqnarray}
The third term in Eq.\,\eqref{eq:effective} describes the interaction between the bosonic mode $b$ and the two-level system induced by the auxiliary mode, and its coupling strength $\chi=gJ(\Delta_{b}^{-1}+\eta_{b}^{-1}+\Delta_{q}^{-1}+\eta_{q}^{-1})/2$ with $\eta_{b(q)}=\omega_{a}+\omega_{b(q)}$ depends on the original system parameters. The last term $\propto (b+b^{\dagger})^{2}$ with coefficient $\xi=J^{2}(\Delta_{b}^{-1}+\eta_{b}^{-1})/2$ induces the squeezing effect of the bosonic mode $b$. The higher-order terms of this unitary transformation can be ignored safely, when the large detuning condition is satisfied. By applying a squeezing operator $S_{b}(r)=\text{exp}[\frac{r}{2}((b ^{\dagger})^{2}-b^{2})]$ with $r=-(1/4)\text{ln}\left(1-4\xi/\omega_{b}\right)$ into the effective Hamiltonian $H_{\text{eff}}$, we obtain
\begin{eqnarray}\label{eq:Rabi}
H_{r}&=&S^{\dagger}_{b}(r)H_{\rm eff}S_{b}(r)\nonumber
\\
&=&\frac{\omega_{q}}{2}\sigma_{z}+\omega_{r}b^{\dagger}b-\chi_{r}(b+b^{\dagger})\sigma_{x}+C_{r},
\end{eqnarray}
where $\omega_{r}=\omega_{b}\text{exp}(-2r)$, $\chi_{r}=\chi\text{exp}(r)$, and $C_{r}=(\omega_{b}/2)[\text{exp}(-2r)-1]$. It obviously shows that the effective interaction between these two uncoupled systems, triggered by virtual excitation of the bosonic mode $a$, is equal to a parameter-dependent Rabi model. Considering the stabilization limitation of $\omega_{r}>0$, the hopping interaction strength $J$ should satisfy
$$J<\frac{\sqrt{\omega_{a}\omega_{b}}}{2}\sqrt{1-\omega_{b}^{2}/\omega_{a}^{2}}.$$
\begin{figure}
\includegraphics[width=0.9\columnwidth]{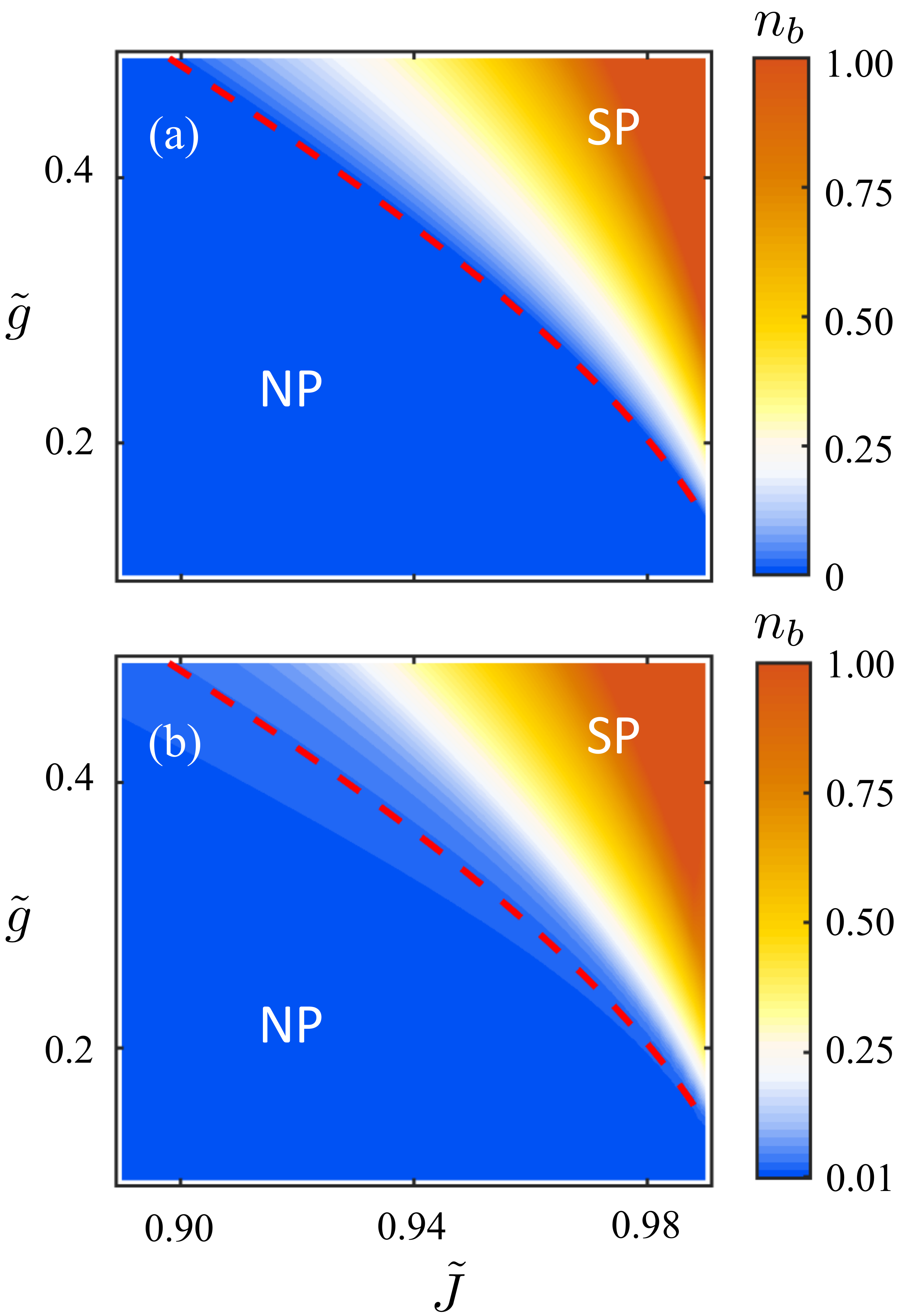}
\caption{\label{fig2} The order parameter $n_{b}$ is plotted as function of the dimensionless parameters $\tilde{J}=2J/\sqrt{\omega_{a}\omega_{b}}$ and $\tilde{g}=2g/\sqrt{\omega_{a}\omega_{q}}$. (a) The analytical result in the $\omega_{q}/\omega_{b}\rightarrow\infty$. (b) The numerical result in finite-parameter ($\omega_{q}/\omega_{b}=5$). Other parameters are $\omega_{a}/\omega_{b}=40$ and $\omega_{b}=1$ GHz. The red dashed contour shows the phase boundary between the normal phase (NP) and superradiant phase (SP).}
\end{figure}

To investigate the ground-state properties, we can diagonalize the Hamiltonian \eqref{eq:Rabi} in the $\omega_{q}/\omega_{b}\rightarrow\infty$ limit (see discussions in Appendix \,\ref{sec B}) and here we have introduced two dimensionless parameters: $\tilde{g}=2g/\sqrt{\omega_{a}\omega_{q}}$ and $\tilde{J}=2J/\sqrt{\omega_{a}\omega_{b}}$. A critical coupling value $\tilde{g}_{c}$ can be derived from the excitation energy $\epsilon_{\text{np}}$ in Eq.\,\eqref{eq:excitation energy}. Specifically, $\epsilon_{\text{np}}$ is real for $\tilde{g}\leq \tilde{g}_{c}$ and vanishes at
\begin{eqnarray}\label{eq:phase boundary1}
\tilde{g}_{c}\approx\frac{\sqrt{1-\tilde{J}^{2}}}{\tilde{J}},
\end{eqnarray}
indicating the occurrence of SPT. The system is in the normal phase for $\tilde{g}<\tilde{g}_{c}$, and it has a ground state $|\phi_{\text{np}}\rangle_{G}=S_{b}(r+r_{\text{np}})|0\rangle_{b}|\!\!\!\!\downarrow\rangle$ with $r_{\text{np}}=-\frac{1}{4}\ln\left[ 1-(4\chi^{2}/\omega_{q})/(\omega_{b}-4\xi) \right]$. In this phase, $\vert\phi_{\text{np}}\rangle_{G}$ has even parity, confirmed by the total excitation number 0. When $\tilde{g}>\tilde{g}_{c}$, the system transitions to the superradiant phase, in which the mode $b$ is macroscopically populated. Whereas the mean occupation number of the ancillary mode $a$ is very small due to the fact that there are no energy exchanges between the ancillary mode $a$ and the two subsystems under large detuning conditions. Now the excitation energy $\epsilon_{sp}$ in Eq.\,\eqref{eq:excitation energy2} is real for $\tilde{g}>\tilde{g}_{c}$ and the ground state becomes double degenerate given by $\vert\phi_{\text{sp}}\rangle_{G}^{\pm}$  (the specific forms shown in the Appendix \,\ref{sec B}). The superradiant phase breaks the $Z_{\textrm{2}}$ symmetry spontaneously, as it is evident from the nonzero coherence of mode $b$, i.e., $\langle b\rangle_{g}=\pm\text{exp}(r)\alpha$.

\begin{figure}
\includegraphics[width=1.0\columnwidth]{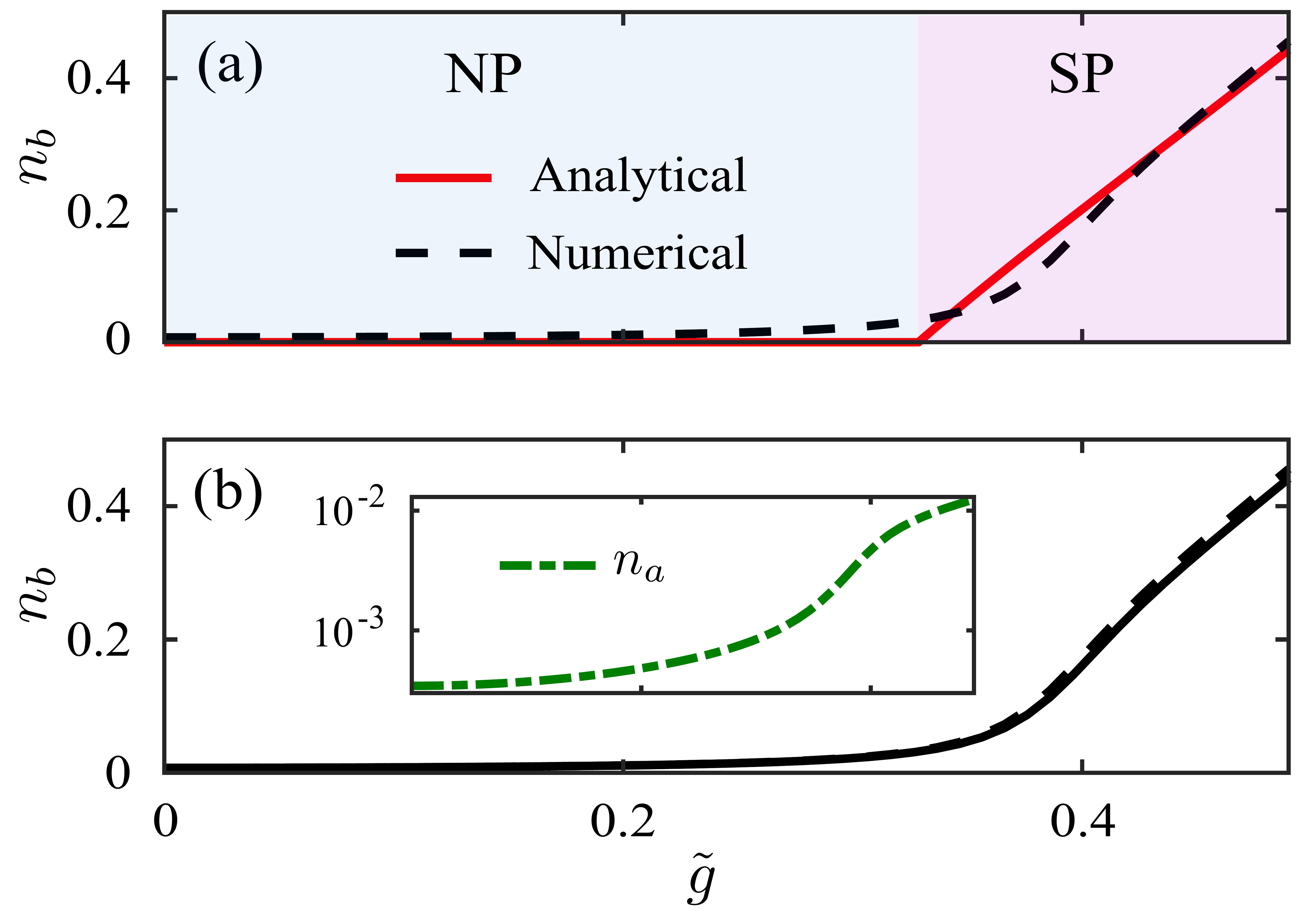}
\caption{\label{fig3} The order parameter $n_{b}$ versus the dimensionless parameter $\tilde{g}$ for $\tilde{J}=0.95$. (a) The analytical result (the red solid contour) and the numerical result (the black dashed line) based on the effective Hamiltonian $H_{\text{eff}}$ when $\omega_{q}/\omega_{b}=5$. (b) The comparison of $n_{b}$ ($n_{a}=[\text{exp}(-4r)\omega_{b}/\omega_{q}]\langle a^{\dagger}a\rangle_{g}$ in the inset) obtained numerically with the initial Hamiltonian in Eq.\,\eqref{eq:initial} (the black solid curve) and the effective Hamiltonian $H_{\text{eff}}$ (the black dashed curve). Other parameters are the same as in Fig.\,\ref{fig2}.}
\end{figure}

The rescaled occupation number $n_{b}=[\text{exp}(-4r)\omega_{b}/\omega_{q}]\langle b^{\dagger}b\rangle_{g}$ is
\begin{eqnarray}
n_{b} & \approx &
\begin{cases}
\tilde{g}^{2}\tilde{J}^{2}/(4\!-\!4\tilde{J}^{2})\!-\!(1\!-\!\tilde{J}^{2})/(4\tilde{g}^{2}\tilde{J}^{2}),& \tilde{g}\!>\!\tilde{g}_{c}\\
0,& \tilde{g}\!<\!\tilde{g}_{c}\\
\end{cases},
\end{eqnarray}
which can be served as an order parameter. In Fig.\,\ref{fig2}, we show the dependence of rescaled excitation number $n_{b}$ on the coupling strengths $\tilde{J}$ and $\tilde{g}$ with the approximate analytic result and the numerical result based on the effective Hamiltonian in finite parameters. The red dashed contour given by Eq.\,\eqref{eq:phase boundary1} indicates the phase boundary that separates the normal phase (NP) $n_{b}=0$ from the superradiant phase (SP) $n_{b}>0$. First of all, one can find that the SPT occurs from the NP to SP by increasing the values of $\tilde{g}$ or $\tilde{J}$. This demonstrates that even though the bosonic mode $b$ do not interact with the two-level emitter directly, this SPT still occurs by introducing an auxiliary mode $a$. The physical mechanism of this SPT is that there exists indirect interaction between these two uncoupled subsystems induced by the virtual excitation number of the auxiliary mode $a$. Moreover, it is also shown from Fig.\,\ref{fig2}(a) that the critical coupling strength $\tilde{g}_{c}$ becomes smaller along with increasing the hopping amplitude $\tilde{J}$. The reason is that both the two coupling channels contribute to the effective interaction $\chi$. Thus, the above results means that here we not only realize the SPT induced by the indirect Rabi interaction, but also obtain a tunable critical point compared with the standard Rabi model whose critical coupling value is fixed at $g_{c_{R}}=\sqrt{\omega_{b}\omega_{q}}/2$ \,\cite{PRL2015Hwang}.  Secondly, comparing the numerical result (Fig.\,\ref{fig2}(b)) with the analytical result, it is shown that the dependence of the order parameter $n_{b}$ on $\tilde{g}$ or $\tilde{J}$ in finite frequency $\omega_{q}/\omega_{b}=5$ approaches to the case of $\omega_{q}/\omega_{b}\rightarrow\infty$. Lastly, Fig.\,\ref{fig2}(b) also shows that the SPT can be observed in a wide coupling range from SC regime ($\{g,J\}/\omega_{a}\gtrsim0.01$) to approximately USC regime ($\{g,J\}/\omega_{a}\gtrsim0.1$), which would loosen the conditions on realizing the SPT in experiment.

In Fig.\,\ref{fig3}, we show the presence of the quantum phase transition more clearly. It is shown from Fig.\,\ref{fig3}(a) that the analytical solution in the $\omega_{q}/\omega_{b}\rightarrow\infty$ limit is consistent with the numerical results based on the effective Hamiltonian \eqref{eq:effective}. Both of them demonstrates that the SPT occurs at the critical point $\tilde{g}_{c}\approx0.3287$, where $n_{b}$ changes suddenly from zero to nonzero. To further show the validity of the above results, we plot the numerical result of $n_{b}$ based on the initial Hamiltonian in Eq.\,\eqref{eq:initial} in Fig.\,\ref{fig3}(b) (the black solid curve), which agrees well with the one obtained by the effective Hamiltonian \eqref{eq:effective} (the black dashed curve). Meanwhile, the inset plots the rescaled occupation of the auxiliary oscillator $a$. In large detuning conditions, $n_{a}$ is smaller than $n_{b}$ by one order of magnitude, which is enough to guarantee the validity of the effective Hamiltonian through adiabatic elimination of the auxiliary mode.

\section{Influence of the $A^2$ term on SPT }\label{sec IV}
\begin{figure}[t]
\includegraphics[width=0.9\columnwidth]{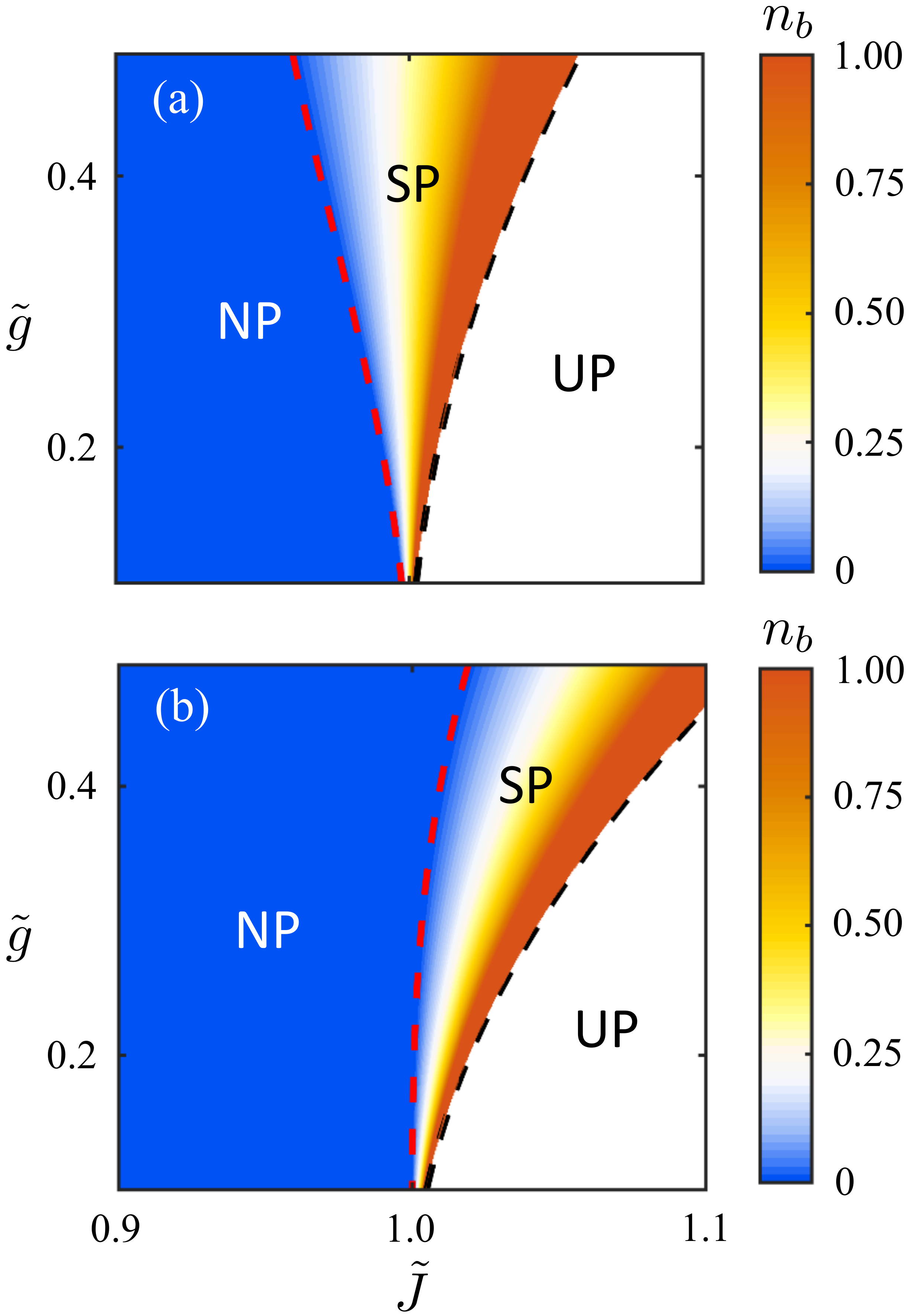}
\caption{\label{fig4}  The order parameter $n^{A}_{b}$ vs $\tilde{J}$ and $\tilde{g}$ for different values of the dimensionless
parameter$\tilde{D}$. (a) $\tilde{D}=0.5$, (b) $\tilde{D}=1.0$. The red (black) dashed contour indicates
the position where $n^{A}_{b}$ becomes nonzero (an imaginary number), locating the phase transition from NP to SP (SP to UP).}
\end{figure}

\begin{figure}[t]
\includegraphics[width=1.0\columnwidth]{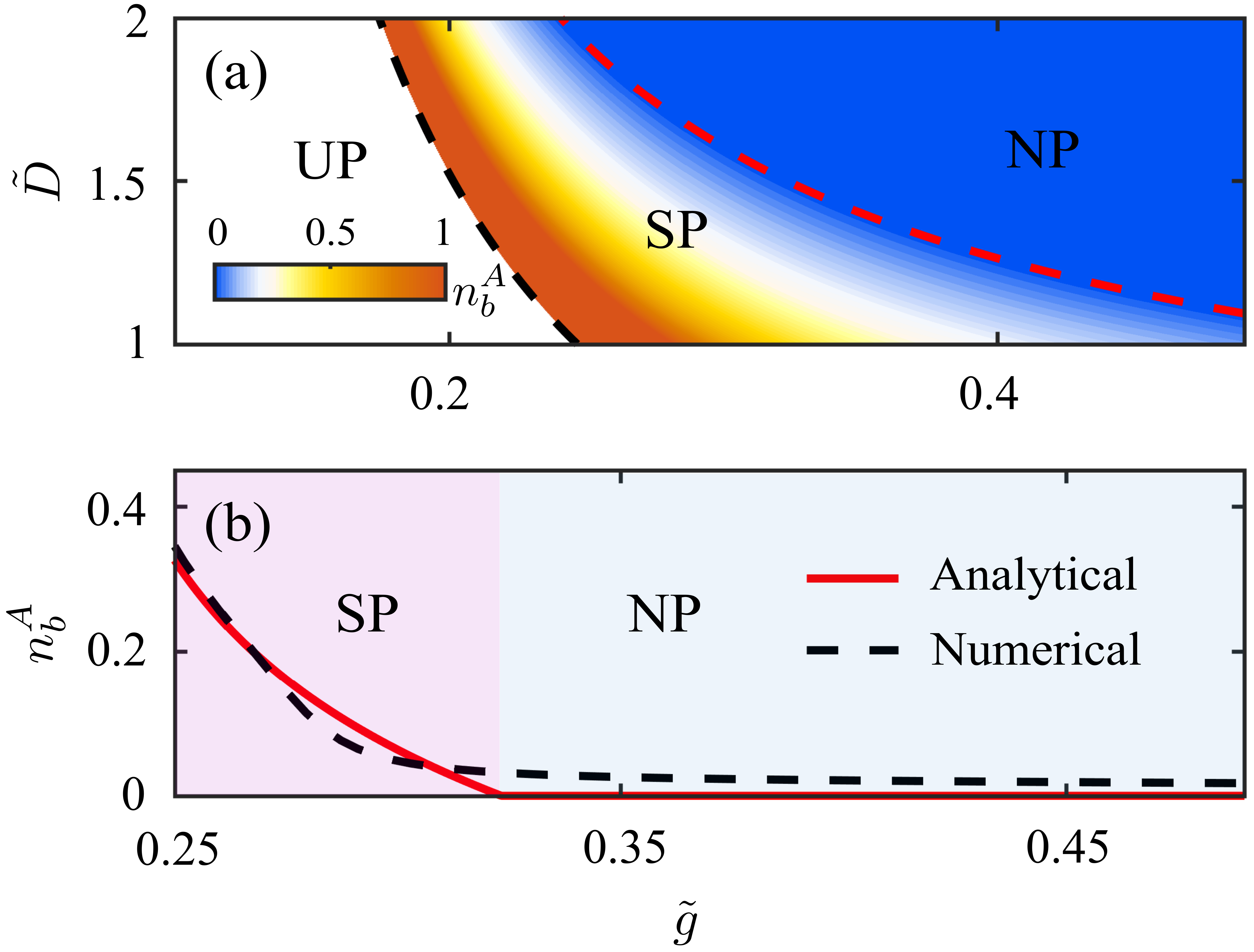}
\caption{\label{fig5}  (a) The order parameter $n^{A}_{b}$ vs $\tilde{g}$ and $\tilde{D}$. The red (black) dashed contour shows the phase boundary between NP and SP (SP and UP). (b) Line cut of the order parameter of $\tilde{D}=1.5$. The analytical result (the red solid contour) and the numerical result (the black dashed line) based
on the effective Hamiltonian Eq.\,\eqref{eq:effA}. We take $\tilde{J}=1.03$ and other parameters are the same as in Fig.\,\ref{fig2}.}
\end{figure}

As is well known, the SPT is challenged by the squared electromagnetic vector potential $A^2$ term, $D(a^{\dagger}+a)^{2}$, of the field stemming from the spin-field interaction. This corresponds to the no-go theorem, which states that the SPT does not occur any longer in the normal Rabi model even if the value of $g$ is very large when $D\geq g^{2}/\omega_{q}$ (determined by the Thomas-Reiche-Kuhn sum rule). We have ignored the $A^2$ term in the previous section, and now we discuss how the $A^2$ term does influence on SPT induced by the indirect spin-field interaction. Incorporating the $A^2$ term into the Hamiltonian $H$ in Eq.\,\eqref{eq:initial} and taking the coefficient $D$ as $\tilde{D}g^2/\omega_q$, the total Hamiltonian $H^{A}=H+\tilde{D}g^2/\omega_q(a^{\dagger}+a)^{2}$. By applying a squeezing transformation $S_{a}(r_{a})=\text{exp}[\frac{r_{a}}{2}((a ^{\dagger})^{2}-a^{2})]$ with $r_{a}=-(1/4)\text{ln}\left[ 1+4\tilde{D}g^2/(\omega_{a}\omega_q) \right]$, the Hamiltonian $H^{A}$ can be transformed into the same form of Eq.\,\eqref{eq:initial}
\begin{eqnarray}\label{eq:A square}
\bar{H}^{A} & = & \frac{\omega_{q}}{2}\sigma_{z}+\bar{\omega}_{a}a^{\dagger}a+\omega_{b}b^{\dagger}b+
\bar{g}(a^{\dagger}+a)\sigma_{x}\nonumber \\
   &  &
   +\bar{J}(a^{\dagger}+a)(b^{\dagger}+b),
\end{eqnarray}
where $\bar{\omega}_{a}\!\!=\!
\!\omega_{a}\text{exp}(-2r_{a})$, $\bar{g}$$=$$g\text{exp}(r_{a})$ and $\bar{J}$$=$$J\text{exp}(r_{a})$.
Accordingly, we can derive the critical coupling strengths $\tilde{g}^{A}_{c}$, $\tilde{J}^{A}_{c}$ and order parameter $n^{A}_{b}$ in the similar way.
The critical coupling strengths $\tilde{g}^{A}_{c}$ and $\tilde{J}^{A}_{c}$ approximately satisfy the following equality
\begin{eqnarray}\label{eq:phaseboundary1}
\frac{\tilde{g}\tilde{J}}{1+\tilde{D}\tilde{g}^{2}}\frac{1}{\sqrt{1-\tilde{J}^{2}/(1+\tilde{D}\tilde{g}^{2})}}\equiv1.
\end{eqnarray}
The order parameter $n^{A}_{b}$ is zero when the left-hand side of Eq.\,\eqref{eq:phaseboundary1} is less than 1
, and $n^{A}_{b}$ is nonzero for the left-hand side of Eq.\,\eqref{eq:phaseboundary1} is greater than 1 and is given by
\begin{eqnarray}\label{eq:order}
n^{A}_{b} & \!\approx\! & \frac{\tilde{g}^{2}\tilde{J}^{2}}{4(1\!+\!\tilde{D}\tilde{g}^{2})(1\!+\!\tilde{D}\tilde{g}^{2}\!-\!\tilde{J}^{2})}
\!-\!\frac{(1\!+\!\tilde{D}\tilde{g}^{2})(1\!+\!\tilde{D}\tilde{g}^{2}\!-\!\tilde{J}^{2})}{4\tilde{g}^{2}\tilde{J}^{2}}.\nonumber \\
         &  &
\end{eqnarray}
Taking into account the stabilization limitation $\omega_{r}^{A}\geq0$ (see Eq.\,\eqref{eq:parameters1}), we find that the SP occurs when $\tilde{J}\leq\sqrt{1+\tilde{D}\tilde{g}^{2}}$. The system enters into the unstable phase (UP) when $\tilde{J}>\sqrt{1+\tilde{D}\tilde{g}^{2}}$, i.e., the white areas under the black dashed curves in In Fig.\,\ref{fig4} and Fig.\,\ref{fig5}(a).

In Fig.\,\ref{fig4}, we show how the order parameter $n^{A}_{b}$ varies as a function of the rescaled coupling strengths $\tilde{J}$ and $\tilde{g}$ for different values of $\tilde{D}$. Fig.\,\ref{fig4}(a) corresponds to the case of $\tilde{D}=0.5$, one finds that system transition from NP to SP by increasing $\tilde{g}$, and the critical coupling strength $\tilde{g}^{A}_{c}$ becomes smaller upon increasing $\tilde{J}$ until a threshold value $\tilde{J}^{A}_{c}\approx0.9978$ determined by Eq.\,\eqref{eq:phaseboundary1}. When $0.9978<\tilde{J}\leq\sqrt{1+\tilde{D}\tilde{g}^{2}}$, the system is bounded in SP within the parameter range. For $\tilde{D}=1$, the system cannot reach the SP even at larger $\tilde{g}$ (corresponding to the no-go theorem) when $\tilde{J}<1$ as shown in Fig.\,\ref{fig4}(b). It is interesting that the SPT recovers when $\tilde{J}>1$ but, conversely, the system enters into the SP by decreasing $\tilde{g}$, which shows a reversed SPT. For $\tilde{D}>1$, the reversed scenario still exists, as shown in Fig.\,\ref{fig5}. One can see that the system enters into the SP when $\tilde{g}$ is smaller than the critical coupling $\tilde{g}^{A}_{c}$ and the critical value decreases with the increased $\tilde{D}$, as shown in Fig.\,\ref{fig5}(a). To display the feature clearly, we choose $\tilde{D}=1.5$ in Fig.\,\ref{fig5}(b). The order parameter $n^{A}_{b}$ suddenly changes from a finite-valued to zero, indicating a reversed SPT. Moreover, the analytical and numerical solutions are still in good agreement. Therefore, the influence of $A^{2}$ term with arbitrary amplitude on SPT can be circumvented when tuning the hopping strength above $\tilde{J}^{A}_{c}$ determined by Eq.\,\eqref{eq:phaseboundary1}.

\section{Schr\"{o}dinger cat state of magnon}\label{sec VI}
\begin{figure}[t]
\includegraphics[width=0.9\columnwidth, clip=true, trim=0 0 0 8]{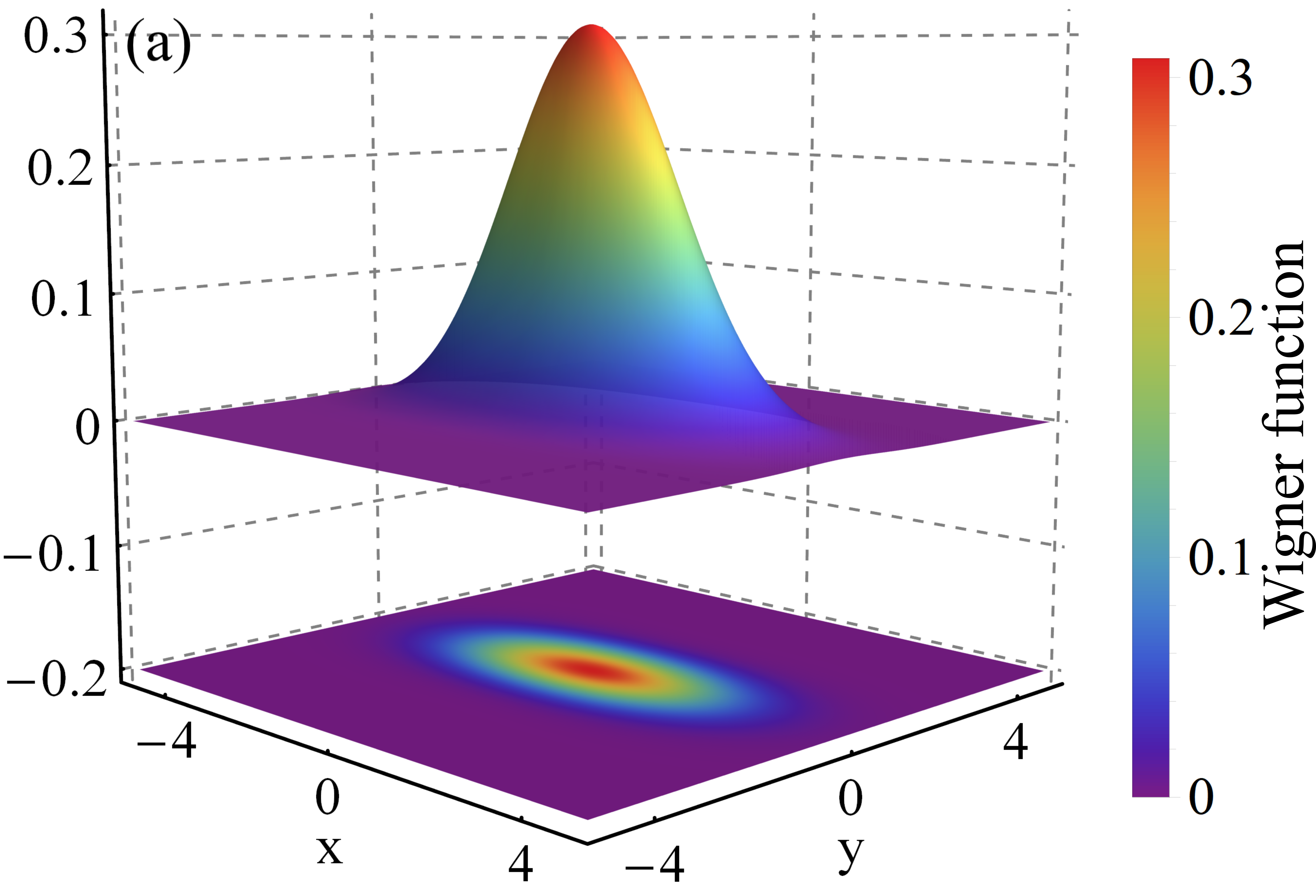}
\includegraphics[width=0.9\columnwidth, clip=true, trim=0 0 0 0]{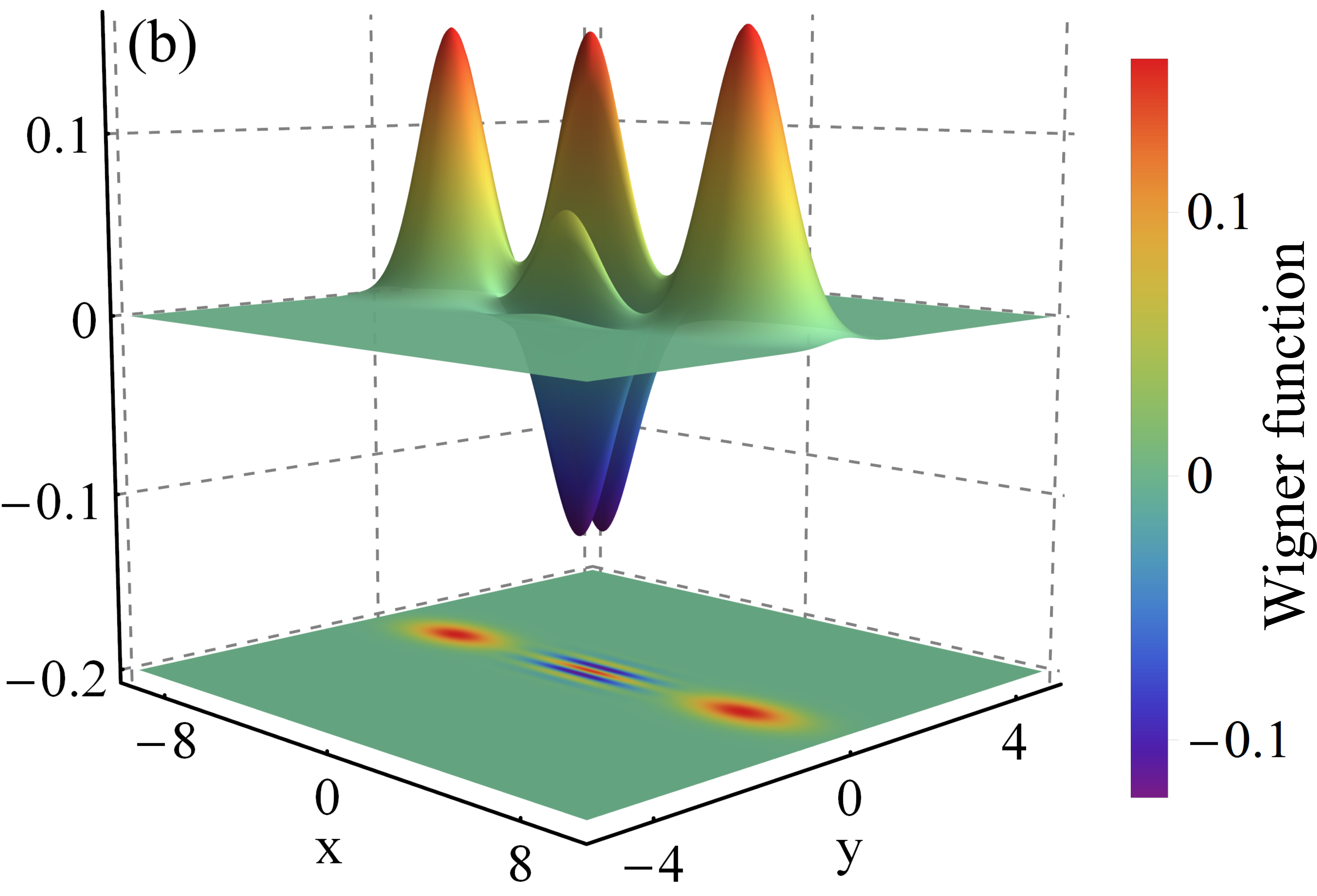}
\caption{\label{fig6} The Wigner functions of the state of oscillator $b$ in the ground state for (a) $g=2.0 \omega_{b}$ (corresponding to the normal phase regime) and (b) $g=3.4 \omega_{b}$ (corresponding to the superradiant phase regime), respectively. The Wigner function distribution in panel (b) displays features of a squeezed cat state (measuring the two-level emitter in $(|\!\downarrow\rangle_{+}+|\!\downarrow\rangle_{-})/\sqrt{2}$). Here, we take $J=3.0 \omega_{b}$ in our numerical calculation and the other system parameters we use are the same as in Fig.\,\ref{fig2}.}
\end{figure}
Considering the implementation of the indirect Rabi model with the hybrid magnon-cavity-qubit system as illustrated in Fig.\,\ref{fig1}(b), our work offers an alternative method for realizing the Schr\"{o}dinger cat state of magnon. Here, the system Hamiltonian can be written in the same form as Eq.\,\eqref{eq:initial}, with $b$ superseded by the magnon operator $m$. As previously discussed, the mediating effect of the microwave cavity field is reflected as an effective coupling between the magnon mode and the qubit, which is the key to indirectly generate a magnon cat state.

To display this, we calculate the Wigner function of the reduced density matrix $\rho_{m}$ based on the effective Hamiltonian\,\eqref{eq:effective}, as plotted in Fig.\,\ref{fig6}. The quadratures are defined as $x=(a+a^{\dagger})/\sqrt{2}$ and $y=(a-a^{\dagger})/\sqrt{2}i$. For moderate coupling strength $g$ (corresponding to the NP), the ground state of the system is a squeezed vacuum state with a Gaussian distribution shown in Fig.\,\ref{fig6}(a). Increasing the coupling strength $g$ above the critical point, the system is in the SP and the ground state can be written as \cite{2010catstate,2013catstate,2020catstate}:
\begin{eqnarray}\label{eq:ground state}
\vert\phi_{\text{sp}}\rangle_{G} & = & \frac{1}{\sqrt2}[S_{m}(r)D(\alpha)S_{m}(r_{\text{sp}})|0\rangle_{m}|\!\downarrow\rangle_{+}\nonumber \\
                              &  &
                              +S_{m}(r)D(-\alpha)S_{m}(r_{\text{sp}})|0\rangle_{m}|\!\downarrow\rangle_{-}],
\end{eqnarray}
where $|\!\!\!\!\!\downarrow\rangle_{\pm}$ is defined in Appendix \,\ref{sec B}. Performing a state measurement on the two-level emitter, the magnon part of the state $|\phi_{\text{sp}}\rangle_{G}$ is projected onto the squeezed cat state $[S_{m}(r)D(\alpha)S_{m}(r_{\text{sp}})\vert0\rangle_{m}\pm S_{m}(r)D(-\alpha)S_{m}(r_{\text{sp}})\vert0\rangle_{m}]/\sqrt{2}$ if the qubit is observed in $(|\!\downarrow\rangle_{+}\pm|\!\downarrow\rangle_{-})/\sqrt{2}$. The properties of the magnon cat state can be characterized by the Wigner function (see Fig.\,\ref{fig6}(b)). The two peaks symmetrical about the $y$ axis are associated with the two cat components $[S_{m}(r)D(\pm\alpha)S_{m}(r_{\text{sp}})\vert0\rangle_{m}$. The other peaks in the center, oscillating between positive and negative values, represent the interference fringes that reveals the coherent superposition of the two components. The system can be prepared in the ground state through adiabatically control the coupling strengths if the change in the control parameters is very slow \cite{nc2021NMR}. Tuning the coupling strengths above the critical value, the system would finally evolve to the cat state. Taking the effect of measurement into account, the system will be inevitably disturbed. Fortunately, the qubit does not directly couple to the magnon mode in our scheme, and thus it is expected to generate a high quality magnon cat state.

To implement this scheme, the key lies in approximately reaching the ultra-strong magnon-to-photon coupling ($J/\omega_{a}\sim0.1$). A directly utilized method is to increase the YIG sphere diameter ($2.5\text{mm}$) and reduce the microwave cavity size ($7.0\times5.0\times3.2\text{mm}^{3}$), which yields a coupling ratio $J/\omega_{a}=6.7\%$ [$J\propto\sqrt{\omega_{a}(V_{m}/V_{a})}$] at room temperature \cite{PRL2014Ultrastrong}, where $V_{a(m)}$ is the volume of the cavity mode (YIG sphere). In addition, this coupling ratio can be further enhanced through novelly cavity engineering and focusing on non-spherical YIG , e.g., placing a YIG block inside a reentrant cavity ($J/\omega_{a}=46\%$ is achieved), or by using a loop gap cavity with a YIG disc (giving $J/\omega_{a}=34\%$) \cite{NJP2019UltrastrongApplications}.

\section{CONCLUSION}\label{sec V}
In this work, we have proposed a scheme to realize the SPT and the associated quantum superposition in a hybrid system described by an indirect Rabi model. This scheme is based on the effective Rabi interaction, between two uncoupled subsystems including the two-level emitter and the bosonic mode $b$, mediated by the virtual excitation of an auxiliary mode $a$. In the large detuning regime and classical oscillator limit $\omega_{q}/\omega_{b}\rightarrow\infty$, we give the analytical results of the ground state both in the NP and SP. In comparison with the standard Rabi model, the critical coupling for SPT becomes adjustable and the SPT still occurs in the presence of $A^{2}$ term. Besides, we show that the cat state can be generated via indirect atom-field interaction. The state of the $b$ mode collapses into a cat state by making projective measurements on the atom, and the resulting cat state may not be disturbed because there is no direct interaction between the atom and the $b$ mode. When our proposal is implemented in the hybrid magnon-cavity-qubit systems, it provides a new method for realizing SPT and macroscopic quantum superposition of magnon \cite{PRB2021spincat,PRL2021magnoncat,PRB2021magnonQPT,CommunicationsPhysics2022MagnonicSPT}, which might expand the cavity magnonics domain and have potential applications in ultra-sensitive magnetic field detection.

\section*{Acknowledgments}
This work is supported by the National Key Research and Development Program of China grant 2021YFA1400700 and the National Science Foundation of China grant No. 11974125.

\appendix
\renewcommand{\appendixname}{APPENDIX}
\section{THE EFFECTIVE HAMILTONIAN IN EQ. (2)}\label{sec A}
\setcounter{equation}{0}
\renewcommand{\theequation}{A\arabic{equation}}
In order to get the effective coupling between the bosonic mode $b$ and two-level system, we apply the Fr{\"o}hlich-Nakajima transformation \cite{Frohlich1950,Nakajima1953}. We now rewrite the total Hamiltonian \,\eqref{eq:initial} into two parts, one is the free Hamiltonian
\begin{eqnarray}
H_{0} & = & \frac{1}{2}\omega_{q}\sigma_{z}+\omega_{a}a^{\dagger}a+\omega_{b}b^{\dagger}b,
\end{eqnarray}
and another one is the interaction Hamiltonian
\begin{equation}
H_{I}=g(a^{\dagger}+a)(\sigma_{+}+\sigma_{-})+J(a^{\dagger}+a)(b^{\dagger}+b).
\end{equation}
We further consider that the bosonic mode $b$ and the two-level system are both coupled to the auxiliary mode $a$ in the large detuning regime, i.e.,
\begin{eqnarray}
\frac{J}{\omega_{a}-\omega_{b}}\ll1, \ \frac{g}{\omega_{a}-\omega_{q}}\ll1.
\end{eqnarray}
Then we can use the Fr{\"o}hlich-Nakajima transformation $U_{V}=e^{V}$ to derive the effective Hamiltonian. The operator $V$ satisfies $H_{I}+[H_{0},V]=0$ and takes the form
\begin{eqnarray}
V & = & \mu_{q}(a\sigma_{-}-a^{\dagger}\sigma_{+})+\nu_{q}(a\sigma_{+}-a^{\dagger}\sigma_{-})\nonumber \\
 &  &
 +\mu_{b}(ab-a^{\dagger}b^{\dagger})+\nu_{b}(ab^{\dagger}-a^{\dagger}b),
\end{eqnarray}
where $\mu_{q}=g/\eta_{q}, \nu_{q}=g/\Delta_{q}$, and $\mu_{b}=J/\eta_{b}, \nu_{b}=J/\Delta_{b}$ and $\eta_{q(b)}=\omega_{a}+\omega_{q(b)}, \Delta_{q(b)}=\omega_{a}-\omega_{q(b)}$. Applying the unitary transformation to $H$ we obtain
\begin{equation}
H_{\text{eff}}=U_{V}^{\dagger}HU_{V}=H_{0}+\frac{1}{2}[H_{I},V]+\frac{1}{3}[[H_{I},V],V]+\cdots.
\end{equation}
Note that the coefficients $\mu_{q(b)}$ and $\nu_{q(b)}$ are small under the large detuning condition, and thus the effective Hamiltonian considered up to second order term $[H_{I},V]$ remains valid with neglecting the higher-order terms. Moreover, we assume that the cavity mode is approximately in the vacuum state. The $b$ mode can couple to the two-level system through virtual excitation of $a$ mode, and thus we eliminate the degrees of freedom of $a$ and obtain
\begin{eqnarray}\label{eq:Heff}
H_{\text{eff}} & = & \frac{\omega_{q}}{2}\sigma_{z}+\omega_{b}b^{\dagger}b-
\chi(b^{\dagger}+b)(\sigma_{+}+\sigma_{-})\nonumber \\
              &  &
              -\xi(b^{\dagger}+b)^{2}+C(a^{\dagger}+a)^{2}+F(a^{\dagger}+a)^{2}\sigma_{z},\nonumber \\
              &  &
\end{eqnarray}
where
\begin{eqnarray}\label{eq:parameters}
\chi & = & \frac{gJ}{2}(\frac{1}{\Delta_{q}}+\frac{1}{\eta_{q}}+\frac{1}{\Delta_{b}}+\frac{1}{\eta_{b}}), \\
\xi & = & \frac{J^{2}}{2}(\frac{1}{\Delta_{b}}+\frac{1}{\eta_{b}}).
\end{eqnarray}
Here, we can ignore the last two terms in Eq.\,\eqref{eq:Heff}, whose coefficients $C=\frac{J^{2}}{2}(\frac{1}{\Delta_{b}}-\frac{1}{\eta_{b}})$ and $F=-\frac{g^{2}}{2}(\frac{1}{\Delta_{q}}-\frac{1}{\eta_{q}})$ are much smaller than both the effective coupling $\chi$ and $\xi$. Therefore, we obtain an effective Hamiltonian
\begin{eqnarray}\label{eq:eff}
H_{\text{eff}}=\frac{\omega_{q}}{2}\sigma_{z}+\omega_{b}b^{\dagger}b-
\chi(b^{\dagger}+b)\sigma_{x}-\xi(b^{\dagger}+b)^{2},
\end{eqnarray}
which is exactly Eq.\,\eqref{eq:effective} of the main text.

Considering the $A^{2}$ term, we can obtain the effective Hamiltonian of the Hamiltonian Eq.\,\eqref{eq:A square} in the same way
\begin{eqnarray}\label{eq:effA}
H_{\text{eff}} & = & \frac{\omega_{q}}{2}\sigma_{z}+\omega_{r}^{A}b^{\dagger}b-
\chi_{r}^{A}(b^{\dagger}+b)\sigma_{x}
\end{eqnarray}
where
\begin{eqnarray}\label{eq:parameters1}
\omega_{r}^{A} & = & \omega_{b}\text{exp}(-2r^{A}), \\
\chi_{r}^{A} & = & \frac{\bar{g}\bar{J}\text{exp}(r^{A})}{2}(\frac{1}{\bar{\Delta}_{q}}+\frac{1}{\bar{\eta}_{q}}+\frac{1}{\bar{\Delta}_{b}}+\frac{1}{\bar{\eta}_{b}}).
\end{eqnarray}
where $\bar{\Delta}_{b(q)}=\bar{\omega}_{a}-\omega_{b(q)}$, $\bar{\eta}_{b(q)}=\bar{\omega}_{a}+\omega_{b(q)}$ and $r^{A}=-(1/4)\text{ln}[1-2\bar{J}^{2}(\bar{\Delta}_{b}^{-1}+\bar{\eta}_{b}^{-1})/\omega_{b}]$.

\section{DETAILS OF DIAGONALIZATION PROCEDURE}\label{sec B}
\setcounter{equation}{0}
\renewcommand{\theequation}{B\arabic{equation}}
In this section, we show details of diagonalizing Hamiltonian\,\eqref{eq:Rabi} in the $\omega_{q}/\omega_{b}\rightarrow\infty$ limit. Concretely, in the NP, performing a Schrieffer-Wolff transformation $U=\text{exp}(S)$ on the Hamiltonian $H_{r}$ in Eq.\,\eqref{eq:Rabi}, we can obtain
\begin{eqnarray}\label{eq:SW}
H_{r}^{\prime}=U^{\dagger}H_{r}U= \sum_{k=0}^{\infty}\frac{[H_{r},S]^{(k)}}{k!},
\end{eqnarray}
where $S=\text{exp}[-i\chi_{r}/\omega_{q}\left(b^{\dagger}+b\right)\sigma_{y}]$, and $[H_{r},S]^{(k)}\equiv[[H_{r},S]^{(k-1)},S]$ with $[H_{r},S]^{(0)}=H_{r}$. Expanding Eq.\,\eqref{eq:SW} up to the second order in $\chi^{\prime}_{r}=2\chi_{r}/\sqrt{\omega_{r}\omega_{q}}$, the transformed Hamiltonian becomes
\begin{eqnarray}
H_{r}^{\prime} & = & \frac{\omega_{q}}{2}\sigma_{z}+\omega_{r}b^{\dagger}b+
\frac{{\chi^{\prime}_{r}}^{2}\omega_{r}}{4}(b^{\dagger}+b)^{2}\sigma_{z}\nonumber \\
              &  &
              +C_{r}+\frac{{\chi^{\prime}_{r}}^{2}\omega_{r}^{2}}{4\omega_{q}}+\mathcal{O}(\frac{{\chi^{\prime}_{r}}^{4}\omega_{r}^{2}}{16\omega_{q}^{2}}).
\end{eqnarray}
In the limit $\omega_{q}/\omega_{r}\rightarrow\infty$, the constant ${\chi^{\prime}_{r}}^{2}\omega_{r}^{2}/(4\omega_{q})$ and high-order terms can be neglected. Considering that the transformed Hamiltonian has spin-up and spin-down subspaces decoupled with each other, the low-energy effective Hamiltonian can be obtained by projecting $H_{r}^{\prime}$ into spin-down subspace, taking the form as
\begin{eqnarray}
H_{\text{np}} & = & \omega_{r}b^{\dagger}b-
\frac{{\chi^{\prime}_{r}}^{2}\omega_{r}}{4}(b^{\dagger}+b)^{2}-\frac{\omega_{q}}{2}+C_{r}.
\end{eqnarray}
Then, we can use a squeezing transformation $S_{b}(r_{\text{np}})=\text{exp}[\frac{r_{\text{np}}}{2}((b^{\dagger})^{2}-b^{2})]$ with $r_{\text{np}}=-(1/4)\text{ln}\left[ 1-4\chi_{r}^{2}/(\omega_{q}\omega_{r}) \right]$, to diagonalize this Hamiltonian in the form as
\begin{eqnarray}
H_{\text{np}} & = & \sqrt{\omega_{r}^{2}-4\chi_{r}^{2}\omega_{r}/\omega_{q}}b^{\dagger}b-\frac{\omega_{q}}{2}+C_{r}\nonumber \\
           &  &
           +\frac{\sqrt{\omega_{r}^{2}-4\chi_{r}^{2}\omega_{r}/\omega_{q}}-\omega_{r}}{2},
\end{eqnarray}
where the ground-state energy is
\begin{eqnarray}
E_{g}=\frac{\sqrt{\omega_{r}^{2}-4\chi_{r}^{2}\omega_{r}/\omega_{q}}-\omega_{r}}{2}-\frac{\omega_{q}}{2}+C_{r},
\end{eqnarray}
and the excitation energy is
\begin{eqnarray}\label{eq:excitation energy}
\epsilon_{\text{np}}=\sqrt{\omega_{r}^{2}-4\chi_{r}^{2}\omega_{r}/\omega_{q}}.
\end{eqnarray}
And the ground state of the system is $\vert\phi_{\text{np}}\rangle_{G}=S_{b}(r+r_{\text{np}})|0\rangle_{b}|\!\downarrow\rangle$ with $S_{b}(r+r_{\text{np}})=\text{exp}[\frac{r+r_{\text{np}}}{2}((b^{\dagger})^{2}-b^{2})]$ possessing even parity, which is confirmed by the total excitation number $0$. A critical coupling value $g_{c}$ can be derived from the vanishing of excitation energy (i.e., $\epsilon_{\text{np}}=0$), and we obtain
\begin{eqnarray}
g_{c}=\frac{\sqrt{\omega_{q}(\omega_{b}-2J^{2}(\Delta_{b}^{-1}+\eta_{b}^{-1}))}}{J(\Delta_{b}^{-1}+\eta_{b}^{-1}+\Delta_{q}^{-1}+\eta_{q}^{-1})}.
\end{eqnarray}
Having introduced two dimensionless parameters: $\tilde{g}=2g/\sqrt{\omega_{a}\omega_{q}}$ and $\tilde{J}=2J/\sqrt{\omega_{a}\omega_{b}}$. The critical condition can be changed into the form independent of frequencies
\begin{eqnarray}\label{eq:phase boundary}
\tilde{g}_{c}\approx\frac{\sqrt{1-\tilde{J}^{2}}}{\tilde{J}}.
\end{eqnarray}
The excitation energy $\epsilon_{\text{np}}$ is real for $\tilde{g}\leq \tilde{g}_{c}$ and zero at $\tilde{g}=\tilde{g}_{c}$, indicating the occurrence of the SPT.

When $\tilde{g}>\tilde{g}_{c}$, the system transitions to the SP, in which the oscillator $b$ is macroscopically populated. Thus, we first apply a displacement operator $D(\alpha)=\text{exp}[\alpha(b^{\dagger}-b)]$  with $\alpha=\pm\sqrt{\frac{\omega_{q}}{4\omega_{r}}\left({\chi^{\prime}_{r}}^{2}-{\chi^{\prime}_{r}}^{-2}\right)}$ upon the Hamiltonian\,\eqref{eq:Rabi} in the main text, and we obtain
\begin{eqnarray}
\tilde{H}_{r} & = & \omega_{r}b^{\dagger}b+\frac{\tilde{\omega}_{q}}{2}\tau_{z}
-\tilde{\chi}_{r}\left(b^{\dagger}+b\right)\tau_{x}
+\omega_{r}\alpha^{2}+C_{r},\nonumber \\
                      &  &
\end{eqnarray}
where the renormalized frequency $\tilde{\omega}_{q}=4\chi_{r}^{2}/\omega_{r}$ and the renormalized coupling strength $\tilde{\chi}_{r}=\frac{1}{4}\omega_{r}\omega_{q}/\chi_{r}$. Here, $\tau_{z}$ and $\tau_{x}$ are the revolved Pauli operators can be expressed in the rotated bases $\left\{ |\tilde{\uparrow}\rangle,|\tilde{\downarrow}\rangle\right\}$ taking forms as
\begin{align}
|\tilde{\uparrow}\rangle &= \text{cos}(\theta)|\!\uparrow\rangle+\text{sin}(\theta)|\!\downarrow\rangle, \\
|\tilde{\downarrow}\rangle &= -\text{sin}(\theta)|\!\uparrow\rangle+\text{cos}(\theta)|\!\downarrow\rangle,
\end{align}
with $\text{tan}\left(2\theta\right)=-4\chi_{r}\alpha/\omega_{q}$. Since $\tilde{H}_{r}$ and the Hamiltonian \,\eqref{eq:Rabi} have the same form, one can diagonalize $\tilde{H}_{r}$ by using the similar procedure described for $H_{\text{np}}$. The diagonalized form of $\tilde{H}_{r}$ is $H_{\text{sp}}=\epsilon_{\text{sp}}{b}^{\dagger}{b}+\tilde{E}_{g}$ with the ground-state energy
\begin{eqnarray}
\tilde{E}_{g} & = &\frac{1}{2}(\sqrt{\omega_{r}^{2}-\omega_{q}^{2}\omega_{r}^{4}\chi_{r}^{-4}/16}-\omega_{r})
-\frac{\chi_{r}^{2}}{\omega_{r}}-\frac{\omega_{q}^{2}\omega_{r}}{16\chi_{r}^{2}}+C_{r},\nonumber \\
                          &  &
\end{eqnarray}
and excitation energy
\begin{eqnarray}\label{eq:excitation energy2}
\epsilon_{\text{sp}}=\sqrt{\omega_{r}^{2}-\frac{\omega_{q}^{2}\omega_{r}^{4}}{16\chi_{r}^{4}}}.
\end{eqnarray}
The ground-state of $H_{\text{sp}}$ is now twofold degenerate given by $|\phi_{\text{sp}}\rangle_{G}^{\pm}=S_{b}(r)D(\pm\alpha)S_{b}(r_{\text{sp}})|0\rangle_{b}|\!\!\downarrow\rangle_{\pm}$ with $r_{\text{sp}}=-(1/4)\text{ln}\left[ 1-\omega_{q}^{2}\omega_{r}^{2}/(16\chi_{r}^{4}) \right]$, and the low-energy states of spin given by
\begin{eqnarray}
|\!\downarrow\rangle_{\pm}\!=\!\sqrt{\frac{1}{2}(1\!+\!\frac{\omega_{r}\omega_{q}}{4\chi_{r}^{2}})}|\!\downarrow\rangle
\pm\sqrt{\frac{1}{2}(1\!-\!\frac{\omega_{r}\omega_{q}}{4\chi_{r}^{2}})}|\!\uparrow\rangle.
\end{eqnarray}
And thus the corresponding $Z_{\textrm{2}}$ symmetry of the ground state is spontaneously broken, as it is evident from the nonzero coherence of oscillator $\langle b\rangle_{g}=\pm\text{exp}(r)\alpha$.

One often use various order parameter to characterize quantum phase transition, and here we have defined the renormalized occupation number of the $b$ mode $n_{b}=[\text{exp}(-4r)\omega_{b}/\omega_{q}]\langle b^{\dagger}b\rangle_{g}$ as an order parameter. We separately calculate $n_{b}$ in the normal phase regime and superradiant phase regime, giving the analytical results
\begin{eqnarray}
n_{b} & = &
\begin{cases}
\frac{\chi^{2}}{\omega_{q}\omega_{b}\!-\!4\omega_{q}\xi}-\frac{\omega_{q}\omega_{b}\!-\!4\omega_{q}\xi}{16\chi^{2}},& g\!>\! g_{c}\\
0,& g\!<\!g_{c}\\
\end{cases}.
\end{eqnarray}
In terms of the rescaled parameters $\tilde{g}$ and $\tilde{J}$, the order parameter approximately becomes
\begin{eqnarray}
n_{b} & \approx &
\begin{cases}
\frac{\tilde{g}^{2}\tilde{J}^{2}}{(4-4\tilde{J}^{2})}-\frac{(1-\tilde{J}^{2})}{4\tilde{g}^{2}\tilde{J}^{2}},& \tilde{g}>\tilde{g}_{c}\\
0,& \tilde{g}<\tilde{g}_{c}\\
\end{cases}.
\end{eqnarray}

\section{THE SPT IN ANISOTROPIC HOPPING INTERACTION}\label{sec C}
\setcounter{equation}{0}
\renewcommand{\theequation}{C\arabic{equation}}
We now consider the case where the hopping interaction is anisotropic, and the Hamiltonian can be rewritten as $H=H_{0}+H_{I}$ in terms of the free Hamiltonian
\begin{eqnarray}
H_{0} & = & \frac{1}{2}\omega_{q}\sigma_{z}+\omega_{a}a^{\dagger}a+\omega_{b}b^{\dagger}b,
\end{eqnarray}
and the interaction Hamiltonian
\begin{equation}
H_{I}=g(a^{\dagger}+a)(\sigma_{+}+\sigma_{-})+J_{1}(ab^{\dagger}+a^{\dagger}b)+J_{2}(ab+a^{\dagger}b^{\dagger}).
\end{equation}
Using the same procedure described in Appendix \,\ref{sec A}, we can derive the effective Hamiltonian. In this case,the Fr{\"o}hlich-Nakajima transformation $U_{V}=e^{V}$ adopts the form
\begin{eqnarray}
V & = & \mu_{q}(a\sigma_{-}-a^{\dagger}\sigma_{+})+\nu_{q}(a\sigma_{+}-a^{\dagger}\sigma_{-})\nonumber \\
 &  &
 +\mu_{b}'(ab-a^{\dagger}b^{\dagger})+\nu_{b}'(ab^{\dagger}-a^{\dagger}b),
\end{eqnarray}
where $\mu_{q}=g/\eta_{q}, \nu_{q}=g/\Delta_{q}$, and $\mu_{b}'=J_{2}/\eta_{b}, \nu_{b}'=J_{1}/\Delta_{b}$ and $\eta_{q(b)}=\omega_{a}+\omega_{q(b)}, \Delta_{q(b)}=\omega_{a}-\omega_{q(b)}$. Then, we eliminate the degrees of freedom of $a$ and obtain
\begin{eqnarray}\label{eq:Heff}
H_{\text{eff}} & = & \frac{\omega_{q}}{2}\sigma_{z}+\omega_{b}b^{\dagger}b-
\chi_{2}(b^{\dagger}\sigma_{+}+b\sigma_{-})-\chi_{1}(b^{\dagger}\sigma_{-}+b\sigma_{+})\nonumber \\
              &  &
              -\xi_{2}(b^{\dagger2}+b^{2})-\xi_{1}(b^{\dagger}b+bb^{\dagger}),\nonumber \\
              &  &
\end{eqnarray}
where
\begin{eqnarray}\label{eq:effective parameter}
\chi_{1} & = & (\frac{gJ_{2}}{(\omega_{a}+\omega_{b})}+\frac{gJ_{1}}{(\omega_{a}-\omega_{b})}
+\frac{gJ_{2}}{(\omega_{a}+\omega_{q})}+\frac{gJ_{1}}{(\omega_{a}-\omega_{q})})/2,\nonumber \\
\chi_{2} & = & (\frac{gJ_{2}}{(\omega_{a}+\omega_{b})}+\frac{gJ_{1}}{(\omega_{a}-\omega_{b})}
+\frac{gJ_{2}}{(\omega_{a}-\omega_{q})}+\frac{gJ_{1}}{(\omega_{a}+\omega_{q})})/2,\nonumber \\
\xi_{1} & = & (\frac{J_{2}^{2}}{(\omega_{a}+\omega_{b})}+\frac{J_{1}^{2}}{(\omega_{a}-\omega_{b})})/2,\nonumber \\
\xi_{2} & = & (\frac{J_{1}J_{2}}{(\omega_{a}-\omega_{b})}+\frac{J_{1}J_{2}}{(\omega_{a}+\omega_{b})})/2.
\end{eqnarray}
\begin{figure}[t]
\includegraphics[width=0.9\columnwidth]{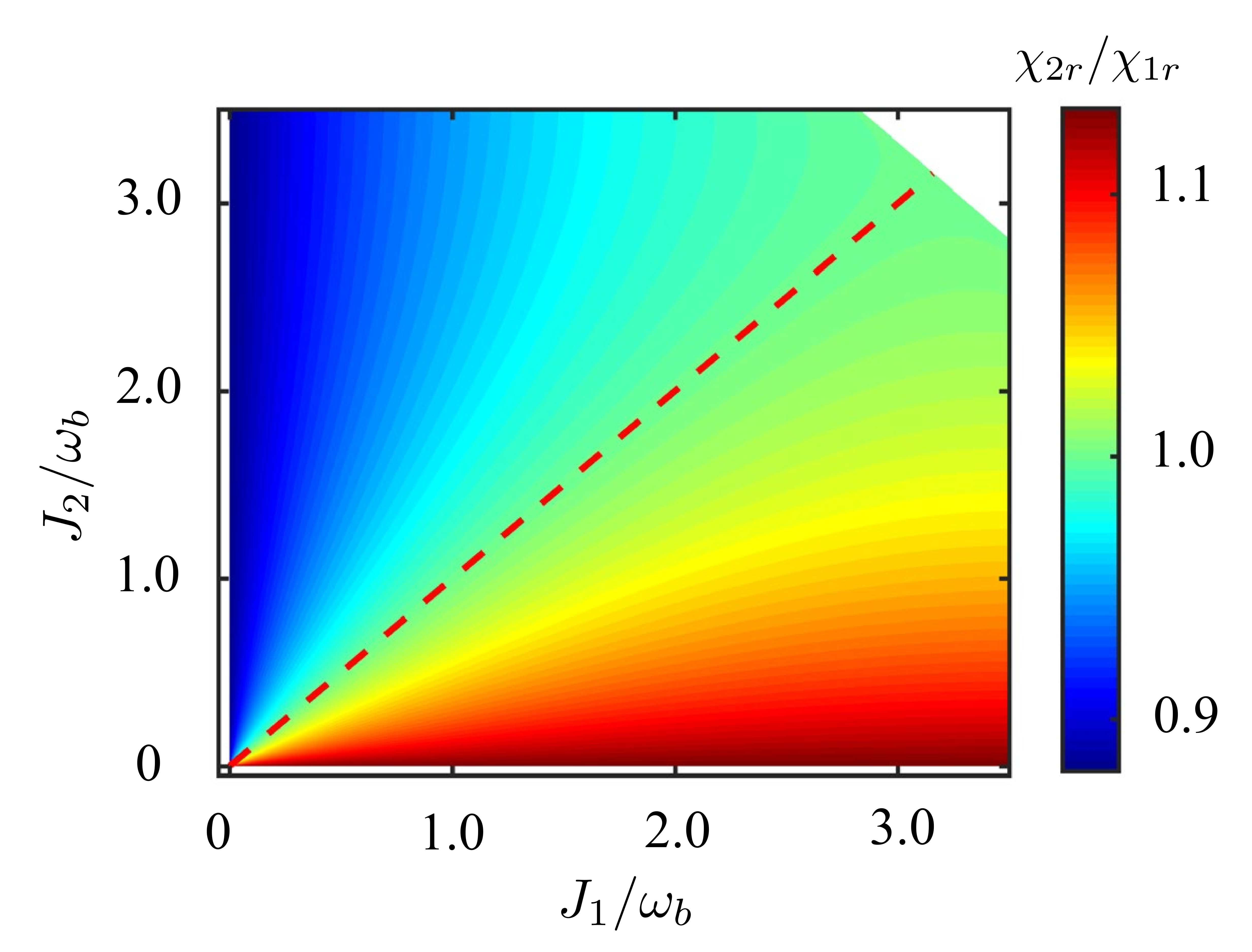}
\caption{\label{fig7} The ratio $\chi_{2r}/\chi_{1r}$ vs $J_{1}/\omega_{b}$ and $J_{2}/\omega_{b}$. The red dashed curve denotes $J_{1}=J_{2}$. We take $g=2.5\omega_{b}$ and other parameters are the same as in Fig.\,\ref{fig2}.}
\end{figure}
Similarly, the effective Hamiltonian remains valid in the large detuning regime, i.e., $g/(\omega_{a}-\omega_{q})\ll1$ and $\{ J_{1},J_{2} \}/(\omega_{a}-\omega_{q})\ll1,\{ J_{1},J_{2} \}/(\omega_{a}-\omega_{b})\ll1$. Different from the effective Hamiltonian Eq.\,\eqref{eq:eff}, this effective Hamiltonian has unequal interaction strengths of rotating-wave and counterrotating-wave terms. By applying a squeezing transformation $S_{b}(r')=\text{exp}[\frac{r'}{2}((b ^{\dagger})^{2}-b^{2})]$ with $r'=-(1/4)\text{ln}\left[1-4\xi_{2}/(\omega_{b}-2\xi_{1}+2\xi_{2})\right]$ into the effective Hamiltonian $H_{\text{eff}}$, we obtain
\begin{eqnarray}\label{eq:AnisotropicRabi}
H_{r'}&=&\frac{\omega_{q}}{2}\sigma_{z}+\omega_{r}'b^{\dagger}b-\chi_{1r}(b^{\dagger}\sigma_{-}+b\sigma_{+})\nonumber
\\
&-&\chi_{2r}(b^{\dagger}\sigma_{+}+b\sigma_{-})+C_{r}',
\end{eqnarray}
where
\begin{eqnarray}\label{eq:parameter}
\omega_{r}' & = & \frac{(\omega_{b}-2\xi_{1})(e^{2r'}\!+\!e^{-2r'})}{2}-\xi_{2}(e^{2r'}\!-\!e^{-2r'}),\nonumber \\
\chi_{1r} & = & \frac{(\chi_{1}+\chi_{2})e^{r}}{2}+\frac{(\chi_{1}-\chi_{2})e^{-r}}{2},\nonumber \\
\chi_{2r} & = & \frac{(\chi_{1}+\chi_{2})e^{r}}{2}-\frac{(\chi_{1}-\chi_{2})e^{-r}}{2},\nonumber \\
C_{r}' & = & \frac{\omega_{b}}{2}(\frac{e^{2r'}\!+\!e^{-2r'}}{2}-1)-\frac{\xi_{1}(e^{2r'}\!+\!e^{-2r'})}{2}\nonumber
\\
       & - &\frac{\xi_{2}(e^{2r'}\!-\!e^{-2r'})}{2}.
\end{eqnarray}
It obviously shows this effective Hamiltonian is equivalent to the anisotropic Rabi model \,\cite{PRL2017anisotropic}.

\begin{figure}
\includegraphics[width=1.0\columnwidth]{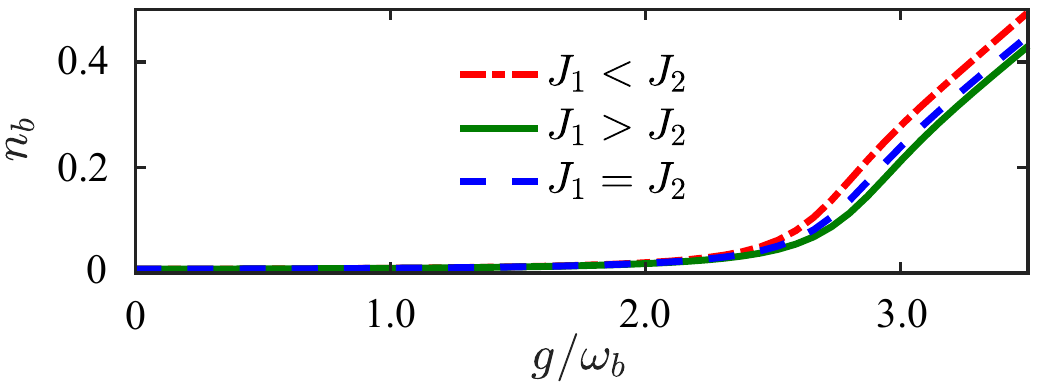}
\caption{\label{fig8} The order parameter $n_{b}$ versus $g/\omega_{b}$ for $J_{1}/\omega_{b}=2.5,J_{2}/\omega_{b}=3.5$ (dash-dotted red curve), $J_{1}/\omega_{b}=3.5,J_{2}/\omega_{b}=2.5$ (solid green curve),$J_{1}/\omega_{b}=3.0,J_{2}/\omega_{b}=3.0$ (dashed blue curve). Other parameters are the same as in Fig.\,\ref{fig2}.}
\end{figure}

In Fig.\,\ref{fig7}, we plot the ratio of $\chi_{2r}$ to $\chi_{1r}$ as a function of coupling strengths $J_{1}$ and $J_{2}$, where the white area in the upper right corner denotes the unstable regime corresponding to the condition $1-4\xi_{2}/(\omega_{b}-2\xi_{1}+2\xi_{2})<0$. It shows that the effective coupling strengths of rotating and counter-rotating terms are approximately close to each other in the large detuning parameter regime. Therefore, the SPT can also exist in the case of $J_{1}\neq J_{2}$, which is same with the case of $J_{1}=J_{2}$ in the main text. To demonstrate this, we show how the rescaled occupation number $n_{b}$ obtained from exact diagonalization varies as a function of the coupling strength $g$ when $J_{1}>J_{2}$, $J_{1}=J_{2}$ and $J_{1}<J_{2}$ in Fig.\,\ref{fig8}. It shows that the behaviors of the order parameter $n_{b}$, changing with the coupling strength $g$, are roughly same for these three cases.
%

\end{document}